\documentclass[traditabstract]{aa}  

\pdfoutput=1

\usepackage{graphicx}
\usepackage{natbib}
\usepackage{amsmath,amssymb}

\begin{document}

\title{ Time-dependent spectral-feature variations of stars displaying the~B[e] phenomenon \thanks{Based on data from the Ond\v{r}ejov 2 m telescope, Czech Republic.} } 
  \subtitle{I. V2028~Cyg}
   \author{ J.~Polster \inst{1,2}
      \and  D.~Kor\v{c}\'{a}kov\'{a} \inst{2}
      \and  V.~Votruba \inst{1,3} 
      \and  P.~\v{S}koda \inst{3} 
      \and  M.~\v{S}lechta \inst{3}
      \and  B.~Ku\v{c}erov\'{a} \inst{1}
      \and  J. Kub\'{a}t  \inst{3}
   }

   \offprints{ J.~Polster, \email{jpolster@email.cz}}
   \institute{Faculty of Science, Masaryk University in Brno, Kotl\'{a}\v{r}sk\'{a} 2, 611 37 Brno, Czech Republic  \\
       \email{jpolster@email.cz}
    \and Astronomical Institute, Charles University in Prague, V Hole\v{s}ovi\v{c}k\'ach 2, 180 00 Praha 8, Czech Republic \\
       \email{kor@sirrah.troja.mff.cuni.cz}
    \and Astronomical Institute of the Academy of Sciences of the Czech Republic,
         Fri\v{c}ova 298, 251 65 Ond\v{r}ejov, Czech Republic}

  \titlerunning{Time-dependent spectral-feature variations of V2028~Cyg}
  \authorrunning{Polster et al.}

  \abstract{

We present results of~nearly six years of~spectroscopic observations of~the~B[e] star V2028~Cyg. The~presence of~the~cold-type absorption lines combined with~a~hot-type spectrum indicate the~binarity of~this object. Since B[e] stars are embedded in~an~extended envelope, the~usage of~common stellar atmosphere models for~the~analysis is quite inappropriate. Therefore, we focus on~the~analysis of~the~long-term spectral line variations in~order to~determine the~nature of~this object. We present the~time dependences of~the~equivalent width and~radial velocities of~the~H$\alpha$ line, [\ion{O}{i}] 6300~\AA, \ion{Fe}{ii} 6427, 6433, and~6456~\AA \, lines. The~bisector variations and~line intensities are shown for~the~H$\alpha$ line. The~radial velocities are also measured for~the~absorption lines of~the~K component. No~periodic variation is found. The~observed data show correlations between the~measured quantities, which can be used in~future modelling.
   }


   \maketitle
%

\section{Introduction}

V2028~Cyg (MWC~623) belongs to~a~rather heterogeneous group called B[e] stars. This group includes such dissimilar objects as compact planetary nebulae, Herbig stars, and supergiants or~symbiotic stars \citep{Lamers}. A~large number of~B[e] stars still belong to~this "unclassified" subgroup, which is also the~case for~V2028~Cyg. It is difficult to~assign this object to~a~specific subgroup because of~the~presence of~a~circumstellar matter. \cite{miros07} found that many unclassified objects display common spectral signatures. He denoted this group as FS~CMa stars, since this star can be taken as a~good prototype for~these objects. He included in~this group V2028~Cyg. Current observation analysis shows that~they are nonsupergiants. FS~CMa objects are probably more evolved stars, but~are either close to the~main sequence, or~still on~it \citep{miros00, miros07}. \cite{miros07} suggested that~a~binarity could explain the~observed properties of~these stars.

The commonly used model of~B[e] stars was proposed by~\cite{zickgraf85}. The~B[e] phenomenon arises in~a~two-component stellar wind with~a~dusty ring. A~slow, cool, high density wind occurs around the~equatorial plane. This wind forms an~outflowing disc, where the~emissions of~both the~Balmer series and~low-ionized metals originate. The~polar areas are~occupied by~a~CAK \citep{CAK} wind of~a~low density and~high velocity. The~gas in~this outflow is hot and~the~emission lines come from~metals in~higher ionization stages.  

V2028~Cyg was first noted as~an~emission-line star by~\cite{merrill42}. \cite{allsw76} discovered its photometrical variability (amplitude 0.5~mag in~filter V). \cite{bergner95} then presented the~amplitude $0.2-0.4$~mag. \cite{allen73} analysed photometric measurements from~the~filters H, K, and~L and~reported an~infrared excess that~he explained in~terms of~a~dust envelope. \cite{allen74} then noted the~connection between~the~excess and~forbidden emission lines. A~few years later, \cite{allsw76} published a~spectral analysis of~the~group of~peculiar Be stars with~infrared excesses, including V2028~Cyg. They identified permitted and~forbidden emission lines, such as~Balmer lines, \ion{Fe}{ii}, [\ion{O}{i}], and~[\ion{Fe}{ii}], which~dominate the~emission spectrum.

V2028~Cyg was noted for~the~first time as~a~binary by~\cite{Arkhipova82}. They suggested spectral the~types B8 and~M1III for~the~system from the~analysis of~spectrophotometric observations. However, they used assumptions that~were later found to~be~incorrect. High resolution spectra enabled \cite{Zickgraf89} to~identify many absorption lines of mainly neutral metals, such as \ion{Fe}{i}, \ion{Ti}{i}, \ion{V}{i}, \ion{Ca}{i}, and~\ion{Li}{i} (also detected later by~\citealt{corporon} and~\citealt{Zickgraf01}). By~fitting theoretical spectral energy distribution to~photometric data, they suggested that~V2028~Cyg is a~binary composed of a~B2 dwarf and~a~K2 giant. From~comparison with~spectra of~normal stars, \cite{Zickgraf01} refined the~component classification to~B4III and~K2Ib-II. Multicolour optical and~near-infrared photometric observations led \cite{bergner95} to~derive the~type K7III for~the~cool component. The~primary component falls on~the~HR diagram into the region where other FS~CMa stars are also found \citep{miros07}. Its position close to~the~region occupied by~classical Be stars, provides a~good opportunity to study the~connection between the~Be and~B[e] phenomenon. 

\cite{Zickgraf01} derived a~kinematical distance towards V2028~Cyg using the~average radial velocity ($2.0^{+0.6}_{-0.3}$~kpc) and~his luminosity estimate for~both components ($2.4^{+1.4}_{-0.9}$~kpc).

The~optical emission spectrum consists of~a~large number of~singly ionized metallic lines (\ion{Fe}{ii}, \ion{Ti}{ii}, \ion{Cr}{ii}, \ion{Si}{ii}) and~is dominated by~the~hydrogen lines of~the~Balmer series. V2028~Cyg possesses an~extended gaseous and~dusty envelope, which is indicated by~several spectral features (IR excess, forbidden lines, intensive Balmer emission). \cite{marston} unsuccessfully tried to~find evidence of~larger-scale structures (lobes, nebula,\ldots) around the~object using a~narrow-band H$\alpha$ filter.

\cite{zicksl89} assumed that~a~circumstellar disc structure causes a~direction-dependent polarisation of~the~emitted photons. They determined an~intrinsic polarization of~$\sim$~2\% from~their observations. On~the~basis of~this value and~the~assumption that~the~scattering plane is~parallel to~the~orbital plane, \cite{Zickgraf01} estimated an~inclination angle $\geq 30-45^{\circ}$. 

The gaseous and dusty envelope of~V2028~Cyg makes it very difficult to~determine the~fundamental parameters of the object, since the~commonly used stellar atmosphere models are not applicable in~this case. The~properties of~these objects can only be described by~multidimensional models that combine time-dependent hydrodynamics and~NLTE radiative transfer. This task is our ultimate goal. Our time-dependent hydrodynamical code \citep{ufo}, which is now extended into 2D, gives the~temperature and~density distributions and~velocity of~the~material for~the~2D radiative transfer code \citep{osa, disk}. However, this is a~long-term project complicated the~role of~different physical effects  in~the~formation of~the~B[e] phenomenon remaining unclear. Therefore, we decided to~focus our efforts here on~the~analysis of~the~object's temporal behaviour that~may reveal important physical mechanisms and~narrow down the~possible system configurations. The~first papers of~the~planned series focus on~a~description of~the~observed spectral properties of~selected B[e] stars over~a~time interval of~several years. 

The~observations, data reduction, and~main spectral features are~described in~Sects. \ref{obs} and~\ref{spec}.  
The following Sect.~\ref{srv} deals with~the~temporal variability of~the~emissions of~\ion{Fe}{ii} (6318, 6456, 6417~\AA), [\ion{O}{i}]~(6300~\AA), and~H$\alpha$. In~Sect. \ref{dis}, we discuss the~possible nature of~V2028~Cyg based on~our observational results and~give several criteria and~suggestions for~future modelling.

\section{\label{obs} Observations and data reduction}

\begin{figure*}[hcbt]
\begin{center}
\centerline{\pdfimage width 18cm {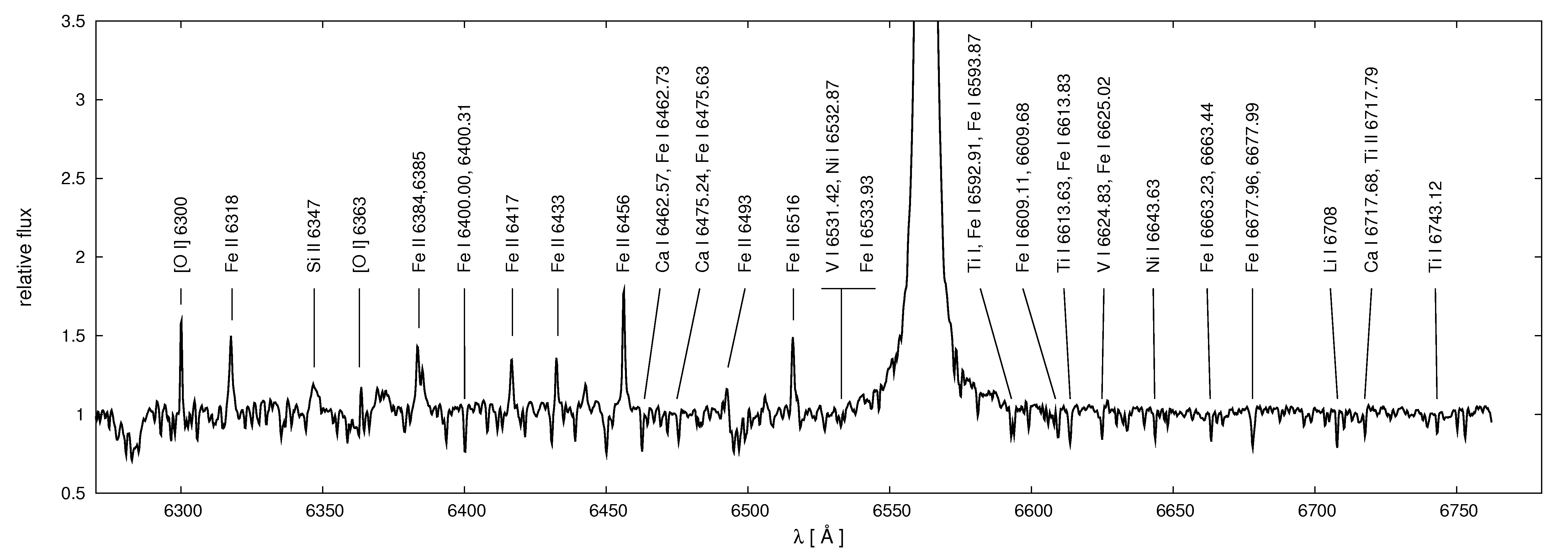}}
\caption{\label{specobr} Spectral interval around H$\alpha$ observed at the Ond\v{r}ejov observatory. H$\alpha$ profile is cut to resolve other spectral lines.}
\end{center}
\end{figure*}

The principal part of~our data were obtained with the~2~m telescope located at~the~Ond\v{r}ejov observatory, Czech Republic. We collected 88 spectra around the~H$\alpha$ line during 45 nights over nearly six years (December~1, 2004 -- October~8, 2010). The~wavelength interval in~these spectra covers $6250-6770$~\AA \,
(Fig.~\ref{specobr}). The~resolving power is R~$\sim 12500$. Three additional spectral intervals were observed: $7520-8030$~\AA \,(seven spectra), $7700-8210$~\AA \,(three spectra), and~$8200-8700$~\AA \,(three spectra). The Ond\v{r}ejov spectra were taken with~the~classical spectrograph at~the~Coud\'{e} focus \citep{skoda02}.

This data set is supplemented by~four spectra (acquired over the~years $1994-1996$) obtained with 
an~\`{e}chelle spectrograph ELODIE operated at~the~Cassegrain focus of~the~1.93~m telescope at~the~Observatoire de~Haute-Provence. The~resolving power is 42000 \citep{elodie} in~the~wavelength interval $4000-6800$~\AA. 

Data reduction was performed using~the~IRAF\footnote{IRAF is distributed by~the~National Optical Astronomy Observatories, operated by~the~Association of~Universities for~Research in~Astronomy, Inc., under contract to~the~National Science Foundation of~the~United States.} data reduction package, and~the~DCR program \citep{pych} was used for~cosmic-ray-hit elimination.

\section{\label{spec} Spectral features}

As~mentioned above, the~spectrum of~V2028~Cyg contains both emission and~absorption lines. The~absorption spectrum is dominated by~the~lines of~the~later component. The~earlier component absorptions are~observed in~\ion{He}{i} lines (4026, 4144, 4472~\AA). The~emission spectrum is dominated by~Balmer lines (H$\alpha$ -- H$\delta$), which are about an~order of~magnitude stronger then the~remaining emission lines in~the~spectrum. Most of~the~emission lines come from~singly ionized metals (\ion{Fe}{ii}, \ion{Ti}{ii}, \ion{Cr}{ii}, \ion{Si}{ii}). Forbidden lines are represented by~[\ion{Fe}{ii}] (4244, 4277, 4287, 7711~\AA) and [\ion{O}{i}] (6300, 6364~\AA). The~emission-line identification is mostly taken from~\cite{Zickgraf89}.

\subsection{Equivalent widths}

The equivalent widths of~the~lines were calculated by~a~numerical integration of~the~line profile using a~trapezoidal rule. The~method of~\cite{Vollmann06} is adopted for~the~error estimation. The~signal-to-noise ratio (S/N) of~individual spectra was calculated by~performing a~linear fitting of~the~continuum around the~studied line. The~weighting function for~the~fit was taken from~\cite{mikulasek03}.

\begin{figure}[hcbt]
\begin{center}
\centerline{\pdfimage width 8cm {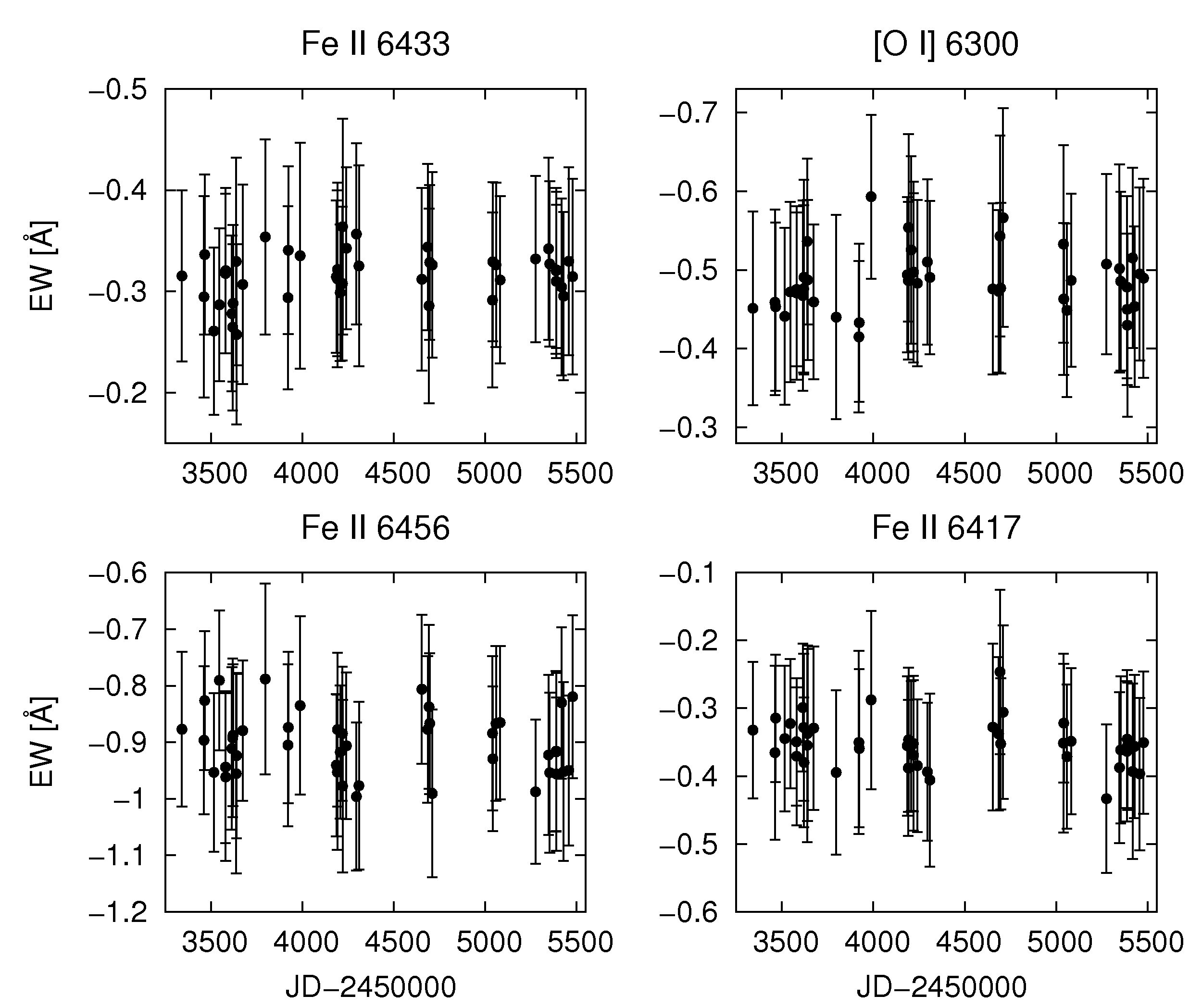}}
\caption{\label{ew}Equivalent widths of~the~\ion{Fe}{ii} and~[\ion{O}{i}] emission lines.}
\end{center}
\end{figure}

The~S/N around the~H$\alpha$ is about~$25-45$. Higher S/N values cannot be achieved using~the~instrument that we used, as~the~maximum intensity of~the~H$\alpha$ line is much stronger than that in the continuum. To~obtain an~unsaturated H$\alpha$ line, the~signal in~the~continuum cannot be high ($300-900$~ADU). The~average spectrum was used when~more then one observation was made during the~night.

\begin{figure}[hcbt]
\begin{center}
\centerline{\pdfimage width 8cm {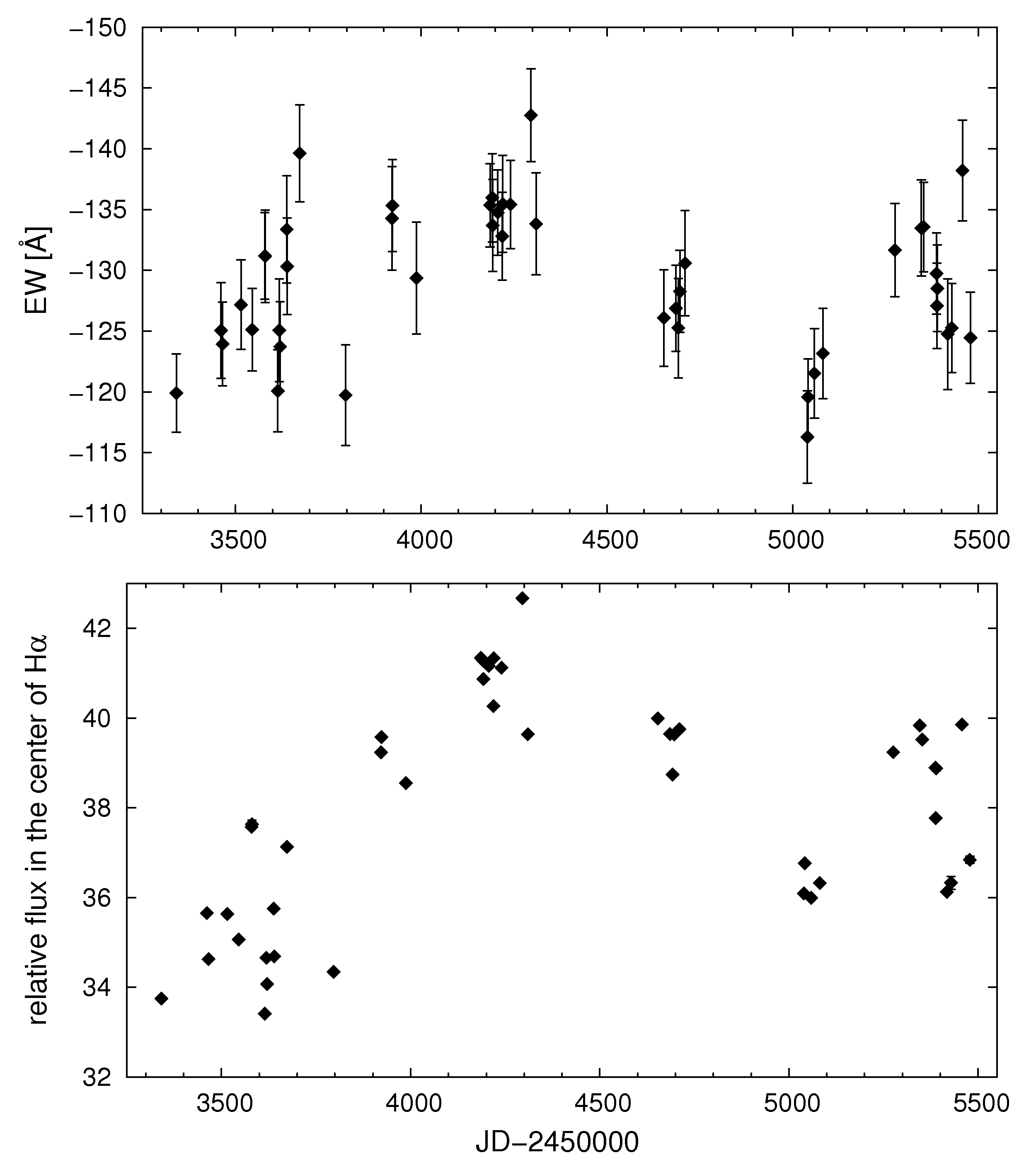}}
\caption{\label{ew_H}Equivalent widths (squares) and~intensities (diamonds) of~the~H$\alpha$ line. Only Ond\v{r}ejov data are presented. The error bars of~intensities are of~the~same size as~the~points.}
\end{center}
\end{figure}

The~equivalent width measurements of~the~metallic lines are shown in~Fig.~\ref{ew}. The~mean values are -0.31$\pm$0.02~\AA (\ion{Fe}{ii} 6433~\AA), -0.90$\pm$0.05~\AA (\ion{Fe}{ii} 6456~\AA), -0.35$\pm$0.03~\AA (\ion{Fe}{ii} 6417~\AA), and~-0.49$\pm$0.04~\AA ([\ion{O}{i}] 6300~\AA). The~relative errors of~$10\% - 26\%$ do not, however, allow us to~make a~definite statement about the~magnitude or~periodicity of~the~variations. 

In~the~case of~the~H$\alpha$ line, the~equivalent width variations are clearly visible. The~relative errors are around 4\%. Fig.~\ref{ew_H} shows the~changes in~the~H$\alpha$ equivalent width and~height. Both quantities exhibit very similar behaviours. Only data from~the~Ond\v{r}ejov spectra are plotted in~Fig.~\ref{ew_H}, therefore direct comparison is possible.

\subsection{Line profiles}

\begin{figure}[!hbct]
\begin{center}
\centerline{\pdfimage width 8cm {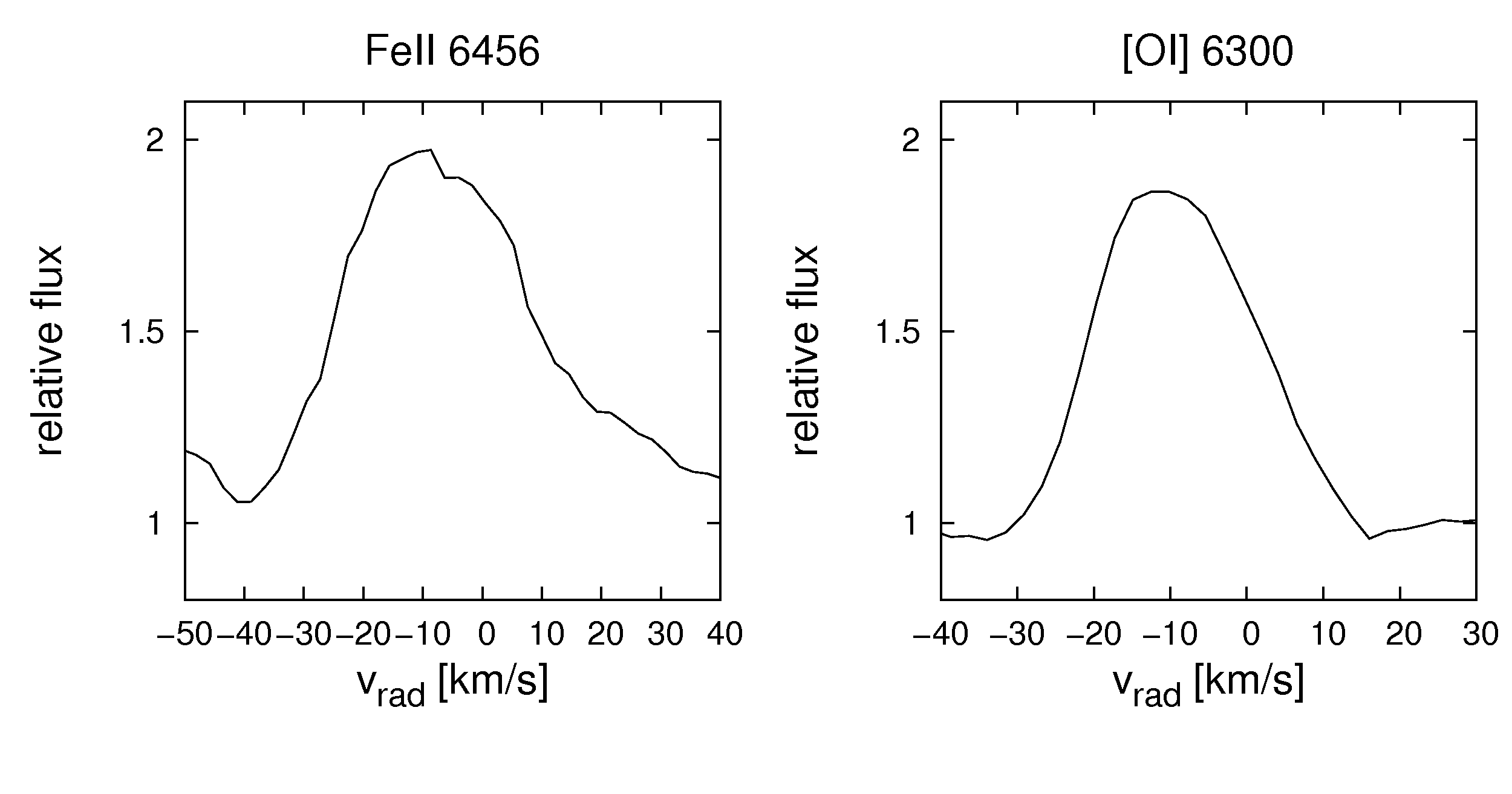}}
\centerline{\pdfimage width 8cm {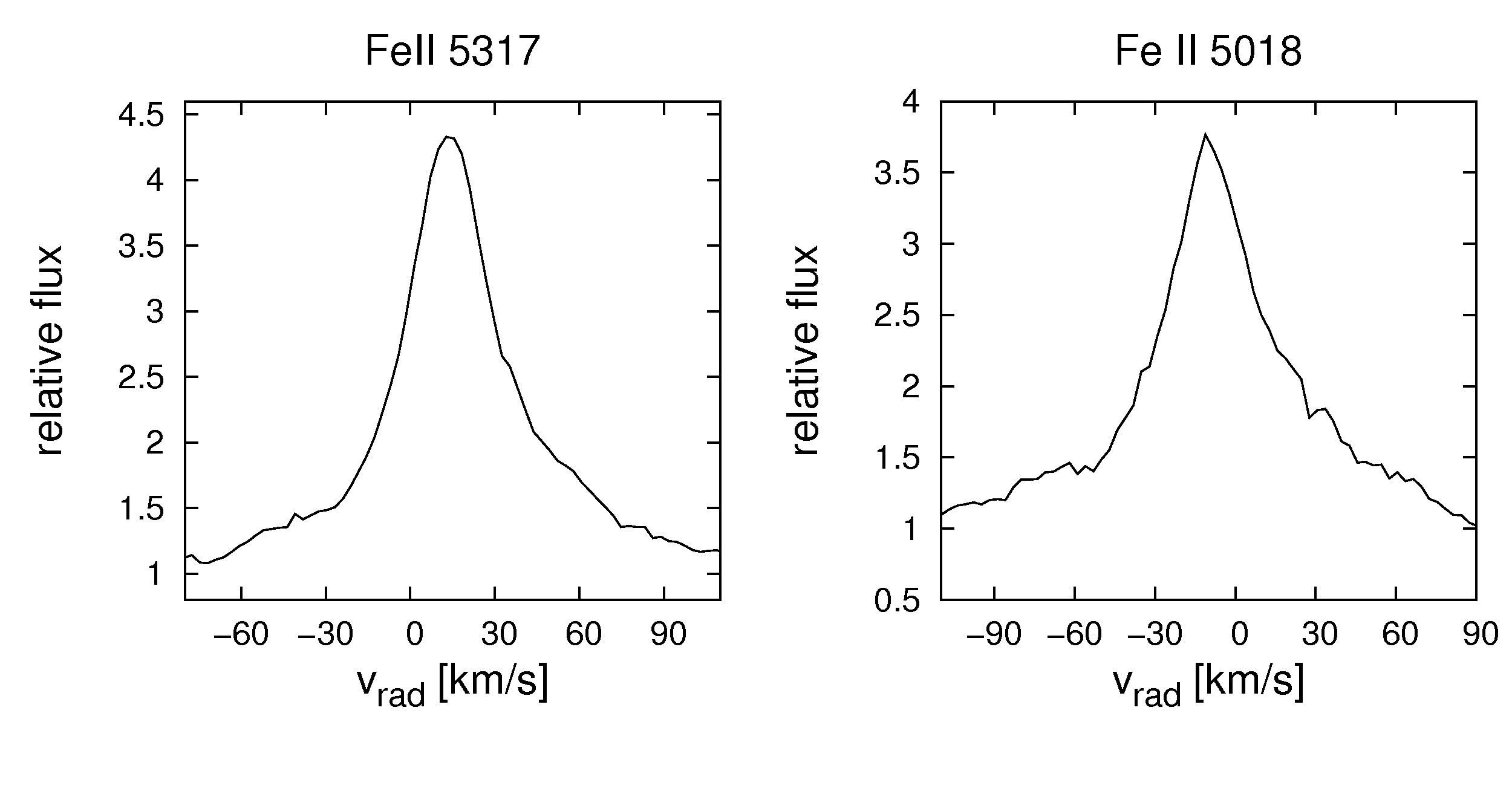}}
\caption{\label{emprof} Emission line profiles of~\ion{Fe}{ii} 6456, 5317, 5018, and [\ion{O}{i}] 6300. ELODIE spectrograph (night August~11/12, 1995).}
\end{center}
\end{figure}

The metallic emission lines in~the~Ond\v{r}ejov spectra are~only slightly asymmetric, and~their profiles are very close to~a~Gaussian. A~higher resolution spectra from the~ELODIE spectrograph clearly show the~asymmetry of~metallic emission lines (see Fig.~\ref{emprof}). 

The~blue side of~the~profile decreases more steeply then the~red side, which is evidence of~a~mass outflow. This feature is common in~all metallic emission lines stronger than $\sim 1.8$ times the~continuum level. The~line profile shape of~the~weaker lines cannot be precisely determined owing to~the~noise, but in~some cases the~same type of~asymmetry is discernible, and~also in~the~case of~the~[\ion{O}{i}] lines. This feature was mentioned before by~\cite{Zickgraf89}.

\begin{figure}[!hbct]
\begin{center}
\centerline{\pdfimage width 8cm {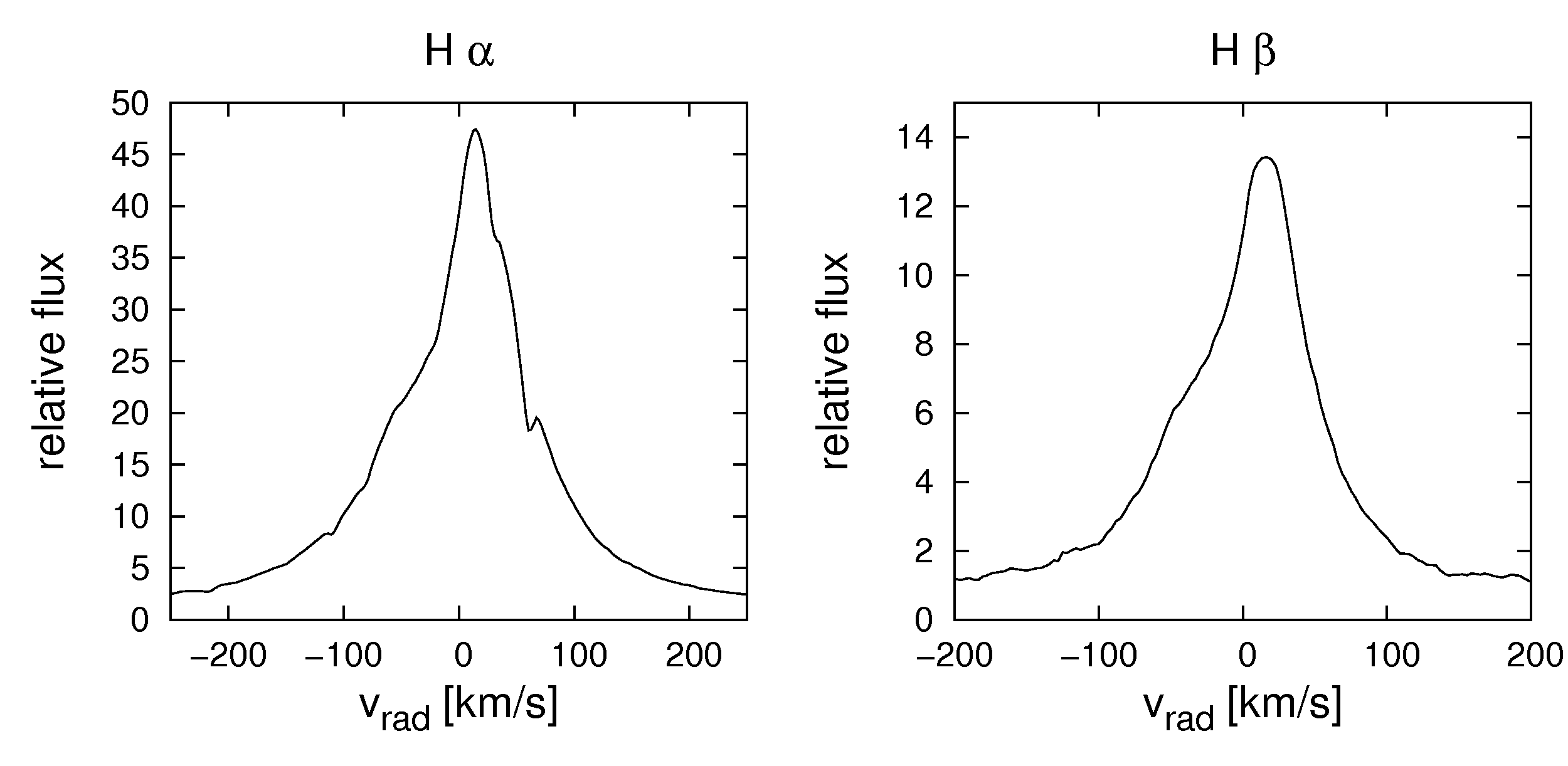}}
\caption{\label{emprofH} Emission line profiles of H$\alpha$ and H$\beta$. ELODIE spectrograph (night August~11/12, 1995). The~line profiles are affected by~the~telluric lines in~this figure.}
\end{center}
\end{figure}

Balmer lines are strongly asymmetric and~their asymmetry is somewhat the~opposite of~that~of~the~metallic lines (Fig.~\ref{emprofH}). The~narrow peak tends to~be shifted to~the~red side with~respect to~the~lower parts of~the~core. The~shift varies in~time. The~little depressions on~the~red side of~the~H$\alpha$ represent the~absorption of~atmospheric water.

The~H$\alpha$ line is the~most pronounced line in~the~spectrum and~its profile is strongly variable. The~changes have~a~noticeable maximum and~minimum and~are seen not only in~the~line intensity and~equivalent width (Fig.~\ref{ew_H}), but~also in~the~bisector velocities. Previous observations of~\cite{Zickgraf01} show even stronger variations then those present in~our data set. He found that the~intensity of~the~H$\alpha$ line changed significantly over the~years $1986-2000$. 

\subsubsection{Width and height of H$\alpha$}

\begin{figure}[!hcbt]
\begin{center}
\centerline{\pdfimage width 8cm {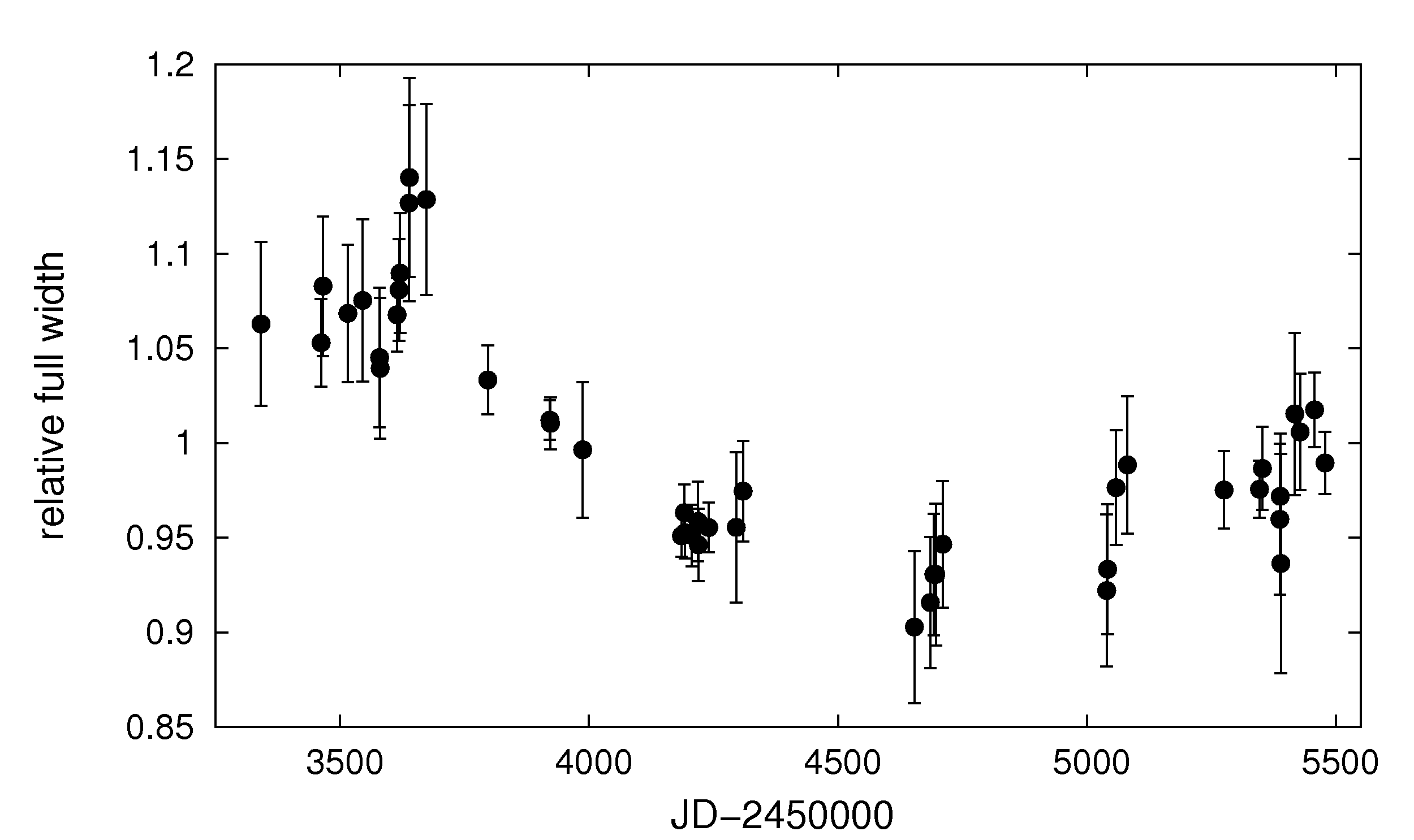}}
\caption{\label{fwp} Variations in~the~full width of~the~H$\alpha$ line. The~mean value of~the~full widths measured in~relative heights $0.1-0.9$ is plotted in~the~graph.}
\end{center}
\end{figure}

The~heights (the~maximum flux $I$; Fig.~\ref{ew_H}) and~widths (Fig.~\ref{fwp}) of~the~H$\alpha$ line vary during the~observation run. The~relative differences of~the~widths are as~high as~24\% and~for~the~heights it is 21\%. 

The~absolute value of~the~equivalent widths $|EW|$ is correlated with height (correlation coefficient $\varrho = 0.75$), but~there is no correlation with~the~measurements of~our so-called full widths ($\varrho = -0.06$), which~were measured at~relative heights of~from~$0.1$ (wings) to~$0.9$ (peak) with a~step size of~0.05. To~compare variations in~the~different relative heights, we divided each of~them by~its mean value. The~relative variations in~the~individual heights were found to~be very similar. 

\subsubsection{H$\alpha$ bisectors}

A~more detailed description of~the~profile variations was done using~bisectors. Certain values of~the~bisectors were measured at~the~relative heights $0.1-0.9$ with a~step of~0.05. We used the~interpolation method of~\cite{steffen90} to~obtain line profile points from~the~observed ones. 

\begin{figure}[hcbt]
\begin{center}
\centerline{\pdfimage width 8cm {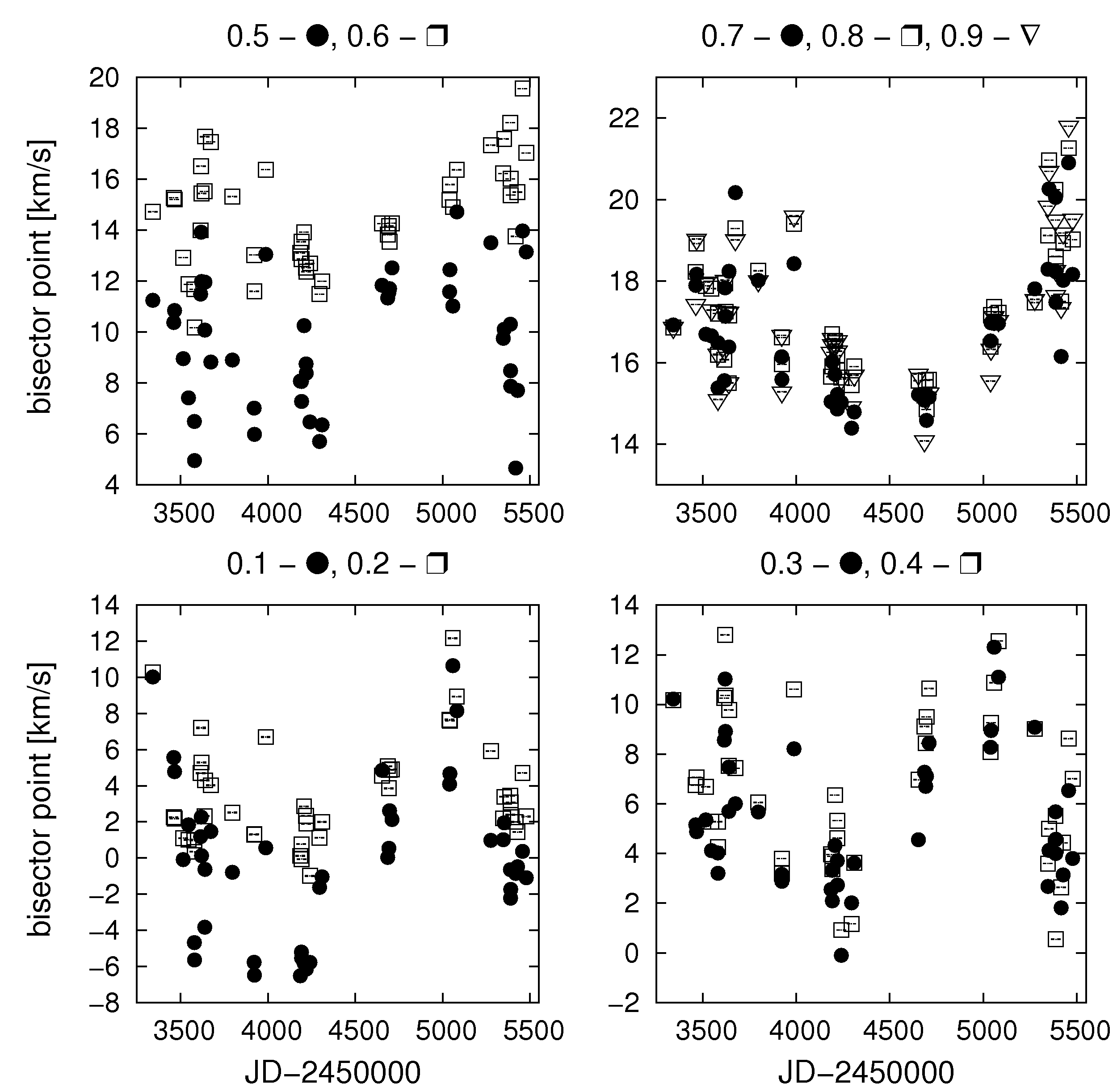}}
\caption{\label{bis_H} Bisector points of the H$\alpha$ line in~relative heights 0.1 (wings) to~0.9 (peak).}
\end{center}
\end{figure}

Three parts of~the~H$\alpha$~line can be resolved (Fig.~\ref{bis_H}). The~wings and~the~lower part of~the~core ($0.1-0.4$ of~the~relative height) appear to~have similar kinematical behaviours and~in~spite of~the~scatter, the~curve with both its minimum and~maximum is discernible in~the~plot. The~minimum occurs on~JD~2454130$\pm$40. The~relative heights at~$0.7-0.9$ represent the~peak. The~bisectors show that~its radial velocity also slightly changes, but in~a~different way from the~velocity variations of~the~wings. The~minimum was observed on~JD~2454540$\pm$30 here. The~bisector velocity at~half-maximum of~the~relative height of~the~line shows some scatter, but~no noticeable trend or~period.

\section{\label{srv} Radial velocities}

Since V2028~Cyg is assumed to~be~a~binary, we investigated the~radial velocity variations in~its spectra. The~absorption lines of~the~cool component as~well as~the~emission from~the~envelope were measured. 

\begin{figure}[hcbt]
\begin{center}
\centerline{\pdfimage width 8cm {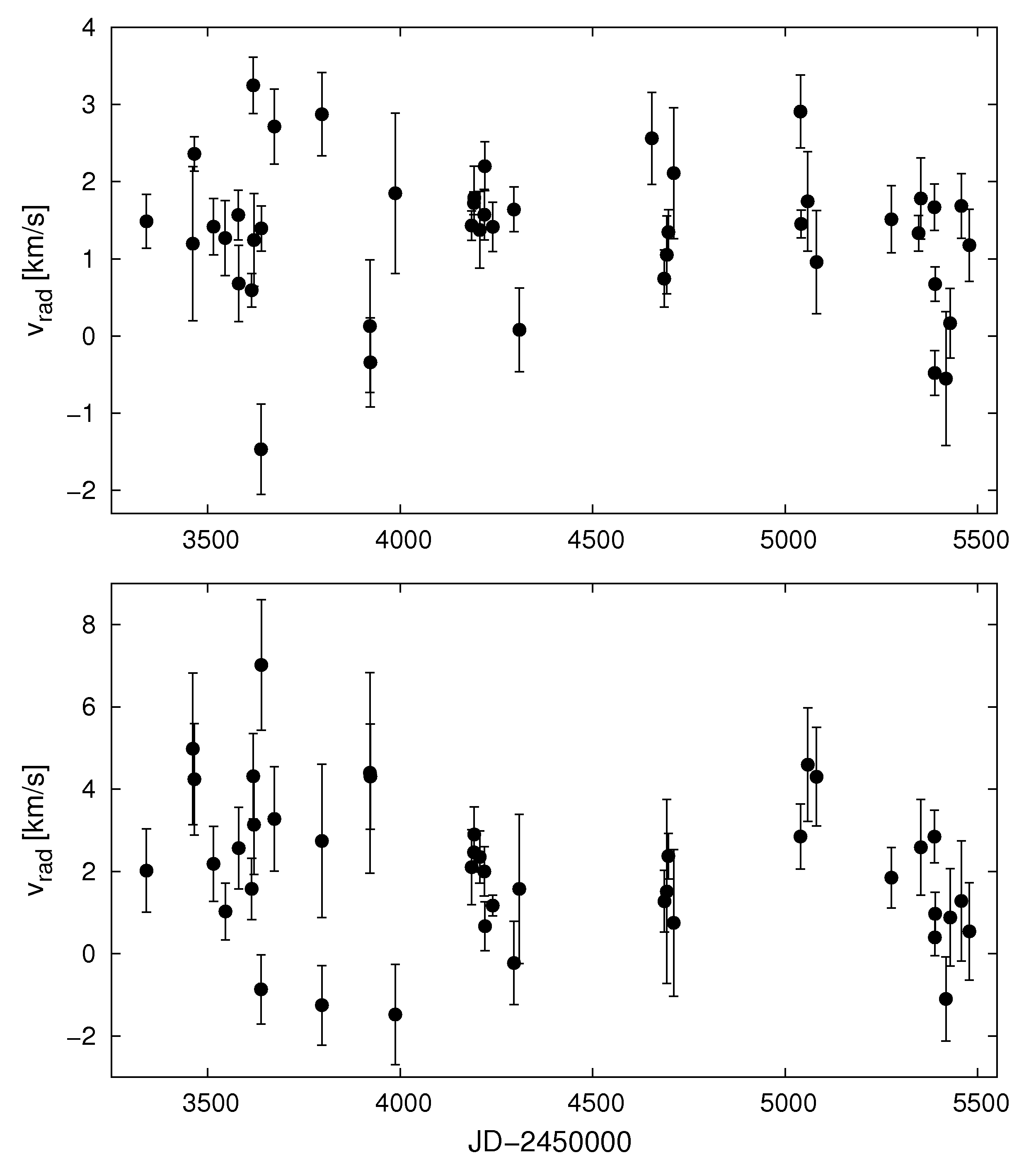}}
\caption{\label{rv_K} Radial velocities of~the~K type absorption lines measured by~cross-correlation (upper panel) and~profile mirroring (lower panel).}
\end{center}
\end{figure}

The~K type absorption lines are often blended and~affected by~noise in~the~Ond\v{r}ejov spectra. We therefore used the~cross-correlation method for~our measurements. The~ELODIE spectrum reduced to the~Ond\v{r}ejov resolution serves as~a~template. Nine intervals without~emission lines and~atmospheric absorption lines were~chosen for~the~measurements  ($6334-6344.5$~\AA, $6398-6415$~\AA, $6434-6440.5$~\AA, $6444-6453$~\AA, $6607-6626$~\AA, $6636-6664.5$~\AA, $6680.5-6693$~\AA, $6694.5-6706.5$~\AA, $6709-6720$~\AA). The~resulting radial velocities in~Fig.~\ref{rv_K} are mean values of~these intervals and~the errors are computed as~the~standard deviation in~the~arithmetic mean. The~measured mean radial velocities vary slightly, but~no~periodical variations were found. To~check the~validity of~the~results we chose seven unblended absorption lines to~measure radial velocities using~other methods (\ion{Fe}{i}~6355~\AA, \ion{Ca}{i}~6439~\AA, \ion{Cr}{i}~6630~\AA, \ion{Fe}{i}~6647~\AA, \ion{Fe}{i}~6648~\AA, \ion{Fe}{i}~6704~\AA, \ion{Li}{i}~6708~\AA). The shifts in~these lines were measured (Fig.~\ref{rv_K}) using~a)~Gaussian fitting and~b)~profile mirroring. The~profile mirroring was done using~both automatic and~manual procedures. The~results from~both techniques agreed to~within the~error intervals. The~lower panel of~Fig.~\ref{rv_K} shows the~values obtained manually using a~program SPEFO\footnote{http://astro.troja.mff.cuni.cz/ftp/hec/SPEFO.263/} \citep[developed by J. Horn, description in][]{skoda96}. The~scatter and~error intervals are~approximately double those of~the~cross-correlation results, although the~variations are very similar. There was a~decline in~the~past 200 days and~had been an~indistinct maximum before it (Fig.~\ref{rv_K}).

\begin{figure}[hcbt]
\begin{center}
\centerline{\pdfimage width 8cm {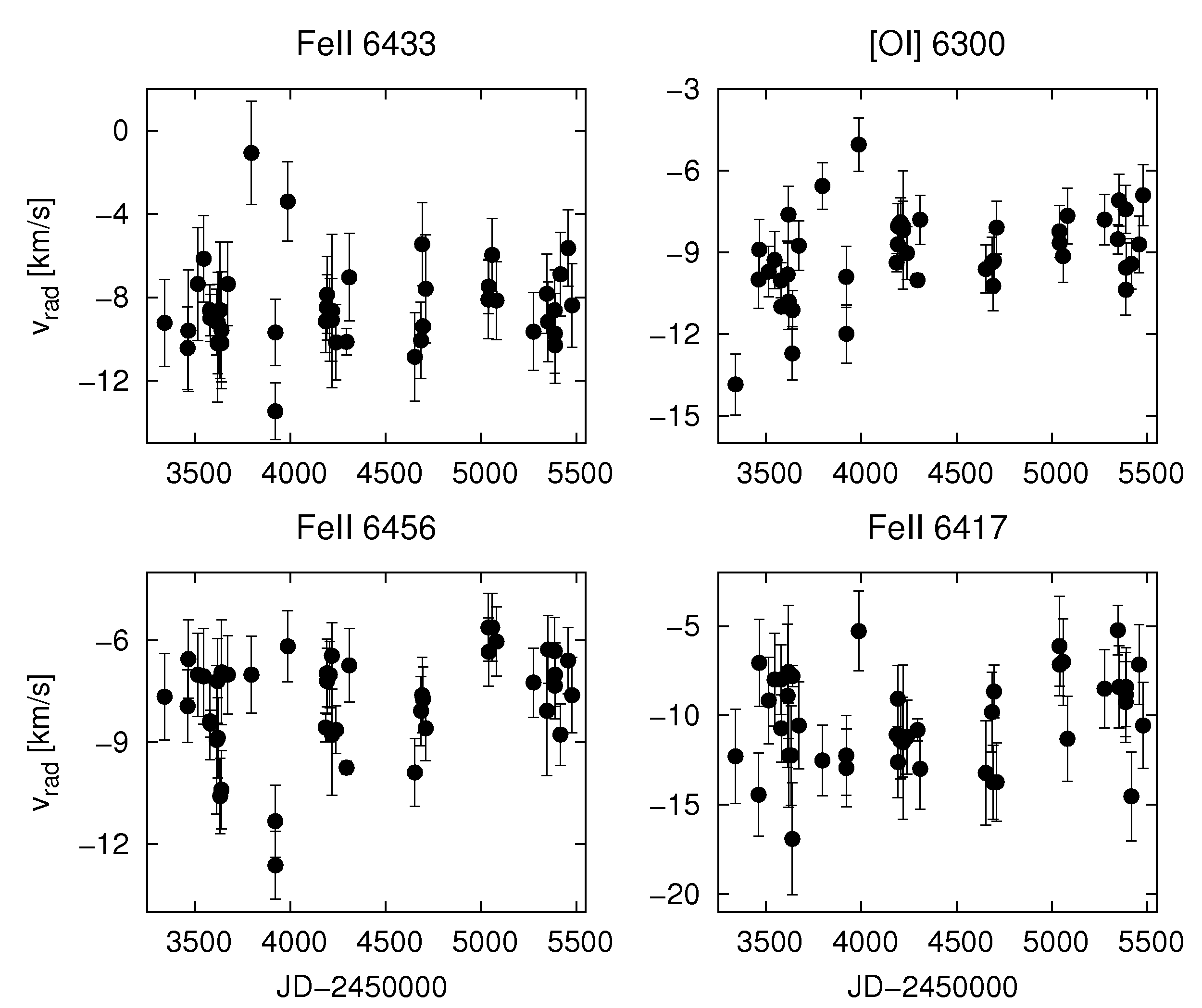}}
\caption{\label{rv_metal} Radial velocities of metallic emission lines.}
\end{center}
\end{figure}

The~radial velocities of~metallic emission lines (Fig.~\ref{rv_metal}) were derived by~Gaussian fitting. There is also some~scatter, but~this is primarily caused by~the~noise. No~period can also be seen here.

The~radial velocities of~the~H$\alpha$ line were measured for~the~peak as~well as~for~the~line wings. A~polynomial fitting was used to~measure the~peak velocities. We adopted the~method of~\cite{mikulasek03} based on~the~least squares method. The~Ond\v{r}ejov spectrograph resolution provides typically six to~eight points in~the~peak interval, which~is sufficient to~obtain a~good fit.

\begin{figure}[hcbt]
\begin{center}
\centerline{\pdfimage width 8cm {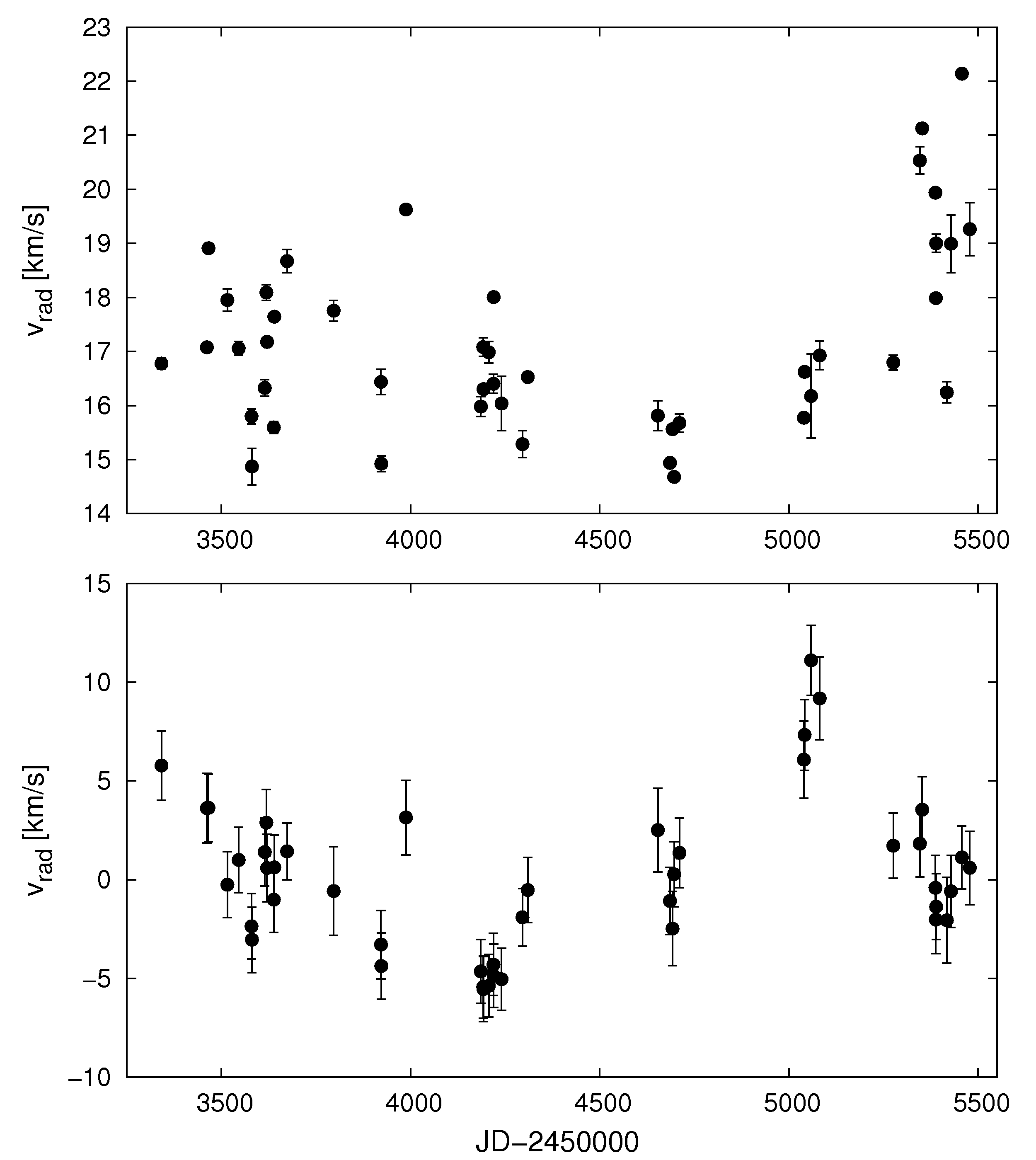}}
\caption{\label{rv_Hpw} Radial velocities of~the~H$\alpha$ peak (upper panel) and~wings (lower panel).}
\end{center}
\end{figure}

We used a~double-Gaussian method \citep{shafter83} to~measure the~radial velocities of~the~wings. The~method consists of~the~determination of~the~line centre $\lambda$ by~solving the~equation

\begin{equation}
\int_{-\infty}^{\infty} S(\Lambda) K (\lambda-\Lambda) \; d \Lambda = 0,
\end{equation}

\noindent
where

\begin{equation}
K(x) = \exp[-(x-a)^{2}/2\sigma^{2}] - \exp[-(x+a)^{2}/2\sigma^{2}]
\end{equation}

\noindent
and~$S(\Lambda)$ is the~investigated spectrum. The~errors were computed according to~the~relation by~\cite{horne86}. The~parameter $a$ adjusts the~separation of~the~Gaussians and~should be~chosen according to~the~width of~the~wings and~the~core. Since the~Gaussian parameter $\sigma$ is limited by~the~spectral resolution, the~lowest possible value in~our case is 0.25~\AA. However, a~larger $\sigma$ value enables us to~measure the~wider wavelength interval of~the~wings and~it deals better with~noise. We adopted values $a = 5.0$~\AA~and~$\sigma = 1.0$~\AA. The~credibility of~this method was verified by~comparing with~the~results obtained by~the~profile mirroring method.

Fig.~\ref{rv_Hpw} (lower panel) shows the~minimum and~maximum of~the~radial velocities of~the~wings. The~position of~the~minimum is JD~2454155$\pm$17. The~radial velocities of~the~H$\alpha$ peak are~plotted in~the~same figure (upper panel). The~results are~different from~those for~the~wing velocities. There is a~large scatter in~the~data but~it is possible to~resolve the~minimum at~JD~2454683$\pm$33.

\section{\label{dis} Discussion and conclusions}

We have measured the~equivalent widths and~radial velocities of~several emission and~absorption lines in~the~spectra of~V2028~Cyg. We have also thoroughly investigated the~profile variations in~the~H$\alpha$ line. 

The~emission lines of~metals (the~most common of~which~is \ion{Fe}{ii}) have a~characteristic bluewards skewed profile, which~is indicative of~an~outflowing envelope. The~line profile width at half-maximum is usually $\sim 45-55$~km/s. The~stronger lines (two times the~continuum intensity) possess wings of~widths up~to~$\sim 200$~km/s. Assuming spherical symmetry, this line profile is indicative of~a~slow (up~to~100~km/s) outflow with a~small velocity dispersion. Equivalent widths and~radial velocities of~the~metallic emissions display some scatter but no~periodic variations. The~slight variations in~the~results may be ascribed to~fluctuations in~the~envelope.

The~forbidden lines are in~the~chosen interval ($\lambda = 6250-6770$~\AA) represented mainly by~[\ion{O}{i}] emission. In~addition, their profiles are skewed bluewards but, however, lack the~broader wings. They therefore probably originate in~the~outer, less dense layers of~the~envelope. 

The~radial velocities of~the~K component measured for~the~single, most clearly defined absorption lines (derived using both the~mirroring method and~by~fitting the~Gaussian profile) have a~similar behaviour to~the~H$\alpha$ wings. However, the~cross-correlation method, even if~the~chosen intervals are~unaffected by~water absorption, gives less significant results. Since our spectra are affected by~a~large amount of~noise, the~radial velocities obtained from~the~measurements of~the~most clearly defined spectral lines are probably more reliable than the~latter. Furthermore, accurate observations are necessary to~(dis)confirm  the~correlation of~the~radial velocities of~the~H$\alpha$ wings with~those of~the~absorption lines. This relation can be important 
when determining the~role of~binarity in~this system. The~radial velocity variations of~the~absorption component in~our data are relatively small ($< 8$~km/s) and~fall into~the~same interval as~the~measurements of~\cite{Zickgraf01}. This implies a~long orbital period for~V2028~Cyg with~a~lower limit of~25~years.

The~Balmer lines H$\alpha$, $\beta$, and~$\gamma$ are the~strongest lines in~the~spectrum (approximately an~order of~magnitude stronger than the~metallic emissions). All of~them are asymmetric with~the~peak shifted redwards with respect to~the~wings. In~most spectra, the~lines bear a~hump on~the~blue side, approximately at~the height of~the~half-maximum. The~H$\delta$ emission is also clearly visible but~any specific features of~the~profile are~overlaid by~the~noise. From this set, only H$\alpha$ has a~sufficient coverage in~the~six-year-long data set from~Ond\v{r}ejov used to~study its variations. It is also a~strongest emission line in~the~spectrum. For~this line, we measured the~variations with~time in~the~equivalent width, intensity, radial velocity, and~bisector values. 

Our observations uncover some of~the~general properties of~the~line profiles and~correlations between measured quantities: 
\begin{itemize}
 \item[--] H$\alpha$ line
  \begin{itemize}
   \item The~line profile is observed in~nearly all spectra to~be~asymmetric. All of~these observations show a~red~shift of~the~line peak relative to~the~centre defined by~the~wings. The~symmetric line profile is very rare in~our~data.
   \item The~radial velocities and~bisectors vary differently in~the~wings and~peak.
   \item The~absolute value of~the~equivalent width $|EW|$ is correlated with~the~maximum flux~$I$ (Fig.~\ref{ew_H}).
   \item The~absolute value of~$|EW|$ is anti-correlated with~the~radial velocity of~the~wings $v_{\rm rw}$ (Fig.~\ref{ew_H} and~\ref{rv_Hpw}).
  \end{itemize}
 \item[--] The~equivalent width of~[\ion{O}{i}] 6300~\AA, \ion{Fe}{ii} 6427, 6433, and~6456~\AA \, lines do not show any significant changes.
 \item[--] The~radial velocities of~the K absorption component may correlate with~the~H$\alpha$ wings. However, further observations of~higher resolution are necessary to~confirm this property. 
\end{itemize}

These results can be compared with~the~results of~future modelling. In~particular, the~anticorrelation of~the~absolute value of~the~H$\alpha$ equivalent width~and~the radial velocities of~its wings probably strongly restricts on~the~proposed model. 

V2028~Cyg is a~very complicated system to~model. It is necessary to~take into~account the~binary nature of~the~object. A~particularly large problem is the~extended circumstellar region, which probably forms a~disc-like structure. These properties require a~combination of~at~least 2D time-dependent hydrodynamic and~NLTE radiative transfer in~moving media. Although our observations have not revealed the~real nature of~V2028~Cyg, they specify the~model properties for~future investigation. Particular model properties can be tested using our observations.

\begin{acknowledgements}
We thank especially Anatoly Miroshnichenko for~his valuable and~inspiring advice and~remarks. We also thank P.~Hadrava, P.~Koubsk\'{y}, and~P.~Harmanec for~their advice and~Adela Kawka for~taking some of~the~spectra in~her observation time. This research is supported by~grants 205/09/P476 (GA \v{C}R), 205/08/H005 (GA \v{C}R), MUNI/A/0968/2009, and~MSM0021620860 (M\v{S}MT \v{C}R). The~Astronomical Institute Ond\v{r}ejov is supported by~a~project AV0Z10030501.
\end{acknowledgements}

\end{document}